# A Planetary Mass and Stellar Radius Relationship for Exoplanets Orbiting Red Giants

Jonathan H. Jiang[1] and Sheldon Zhu[2]

1. Jet Propulsion Laboratory, California Institute of Technology, Pasadena, CA
2. Department of Physics, University of Cambridge, Cambridge, UK.

In about 5 billion years, our sun will leave the main sequence and become a red giant, incinerating most of the inner terrestrial planets, including the Earth [e.g., *Ramirez and Kaltenegger*, 2016]. Once this happens, the outer planets, Jupiter, Saturn, Uranus, and Neptune, would survive in a rearranged planetary system orbiting the red giant sun.

Recent exoplanet discoveries have revealed at least 154 red giant systems in our neighborhood. We collected data from the NASA Exoplanet Archive (https://exoplanetarchive.ipac.caltech.edu/; Archive, here after) for 3451 exoplanets' host stars and plotted them on the Hertzsprung–Russell (HR) diagram, as shown in Figure 1a. In that graph, the x-axis shows $T(K)$, the effective temperature of the host star (in Kelvin) provided in the Archive. The y-axis shows $M_{abs}$, the star's absolute magnitude, computed using formula $M = m - 5\ log_{10}(d_{pc}/10)$, where $m$ is apparent magnitude and $d_{pc}$ is the distance (in parsecs) of the host star; both data are provided in the Archive. The separation of main sequence stars (black dots) and red giants (red dots) is clear in the HR diagram. We checked all the red dots based on their HD numbers and confirmed that they are indeed red giants.

Figure 1b is a scatter plot of exoplanet mass (in units of Earth mass, $M_⊕$) against red giant host star radius (in units of solar radius, $R_⊙$). It demonstrates an interesting positive trend: larger stars have more massive planets. It is important to note that this is a relationship between the stellar radius and the mass of the exoplanets, not between the masses of the star and planet. This implies that the evolution of a star towards a red giant affects the masses of planets in the system. We also examined all the stars available in the Archive and found that this positive trend only exists among the red giants (Figure 1c). Main-sequence stars do not have such clear relationship between planetary mass and stellar radius.

To investigate if the trend is significant, we performed a linear regression. We found a slope of 0.88 with 1-$\sigma$ confidence levels at (0.73, 1.03) and intercept at 2.18 with (2.021, 2.3) bounds for the 1-$\sigma$ confidence level. This mass-radius relation can be written as a power law,

$$\frac{M}{M_⊕} = a \left(\frac{R}{R_⊙}\right)^b \qquad (1)$$

with $b = 0.88$ (0.73, 1.03) and $a = 150$ (104, 214). The $R^2$ value is 0.46, and no systematic bias is apparent in the residuals.

We noted that the trend may potentially be an artifact of detectability drop-off. The majority of exoplanets around red giants were detected through radial velocity (RV) measurements, as transit photometry suffers greatly from the inflated stellar radius, producing extremely small transit depth values. The equation for the semi-amplitude ($K_1$) of radial velocity oscillation of the star is,

$$K_1 = \left(\frac{2\pi G}{P}\right)^{1/3} \frac{m_2 \sin i}{m_1^{2/3}} \frac{1}{\sqrt{1-e^2}} \qquad (2)$$

where $m_1, m_2, P, i,$ and $e$ are the mass of the star, mass of the planet, period, inclination, and eccentricity, respectively [*Paddock* 1913]. Using Kepler's third law and assuming the stellar radius as the minimum semi-major axis, we find the scaling relation,

$$\frac{M}{M_\oplus} \propto K_1 \left(\frac{R}{R_\odot}\right)^{1/2}. \tag{3}$$

Compared to equation (1), this detectability relation scales more slowly, providing a lower bound to the population. To see if this lower bound is applicable, we use limiting $K_1$ values calculated by *Plavchan et al.* [2015] for 10-m telescopes. We also note the fact that most planets will have low eccentricity due to tidal circulation. We assume head-on inclination (90°) and solar mass, and then over-plotted the calculated lower bound as the dashed blue line (Figure 1b). The bound lies well below the majority of the red giant exoplanet population, indicating that detectability is not the main cause of the trend. The trend could also be due to other sensitivity issues, for example, the RV method only yields minimum not absolute masses for the exoplanets. However, the trend does not show up clearly in the planetary mass versus stellar mass plot (Figure 1d).

Current theories of planet formation and evolution support the presence of gas giants orbiting outside the red giant expansion radius but give no mass scaling relation as a function of stellar radius. Hence, we believe that the red giant evolution has an important role in creating the trend itself. A possible explanation may be the increased stellar wind from the inflated star. *Ramirez and Kaltenegger* [2016] give an estimate for the planetary atmospheric mass loss rate. They find that the mass loss rate scales inversely related to the stellar radius, which can produce the positive trend seen. However, due to the lack of evaporation timescales, the process is not well quantified.

The survival of gas giants around red giants is a topic of ongoing investigation. As more exoplanets are found past our current detection limits, we will see whether the trend holds. If so, these red giant exoplanets will help to provide a glimpse into the future of our solar system.

## Acknowledgement

This work was partly supported by the Exoplanet Science Initiative at the Jet Propulsion Laboratory, California Institute of Technology, under contract with NASA. We thank Prof. Yuk Yung, Drs. David Crisp, Renyu Hu and Vijay Natraj for helpful comments.

## Figure Caption

**Figure 1:** (a): The HR diagram of 612 exoplanet host stars. The black-dots are main sequence stars, and the red-colored dots are stars that have been identified as red giants; (b) Scatter plot of exoplanet mass (in units of Earth mass) versus red giant host star radius (in units of solar radius); (c) Scatter plot of planet masses against stellar radii, identifying the red giant planets showing distinctive separation in the main sequence star and red giant populations; (d) Scatter plot of planet masses against stellar masses, also identifying the red giant planets.

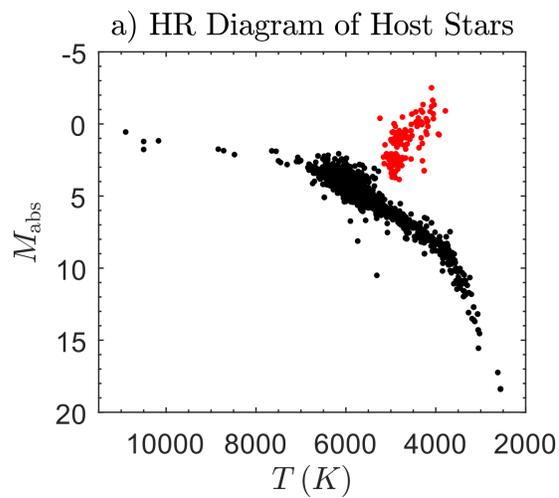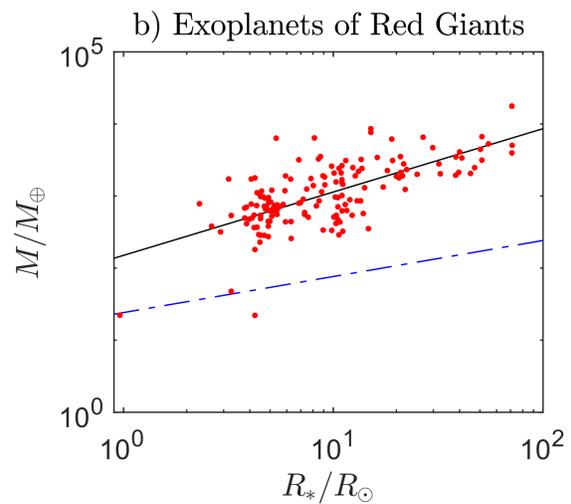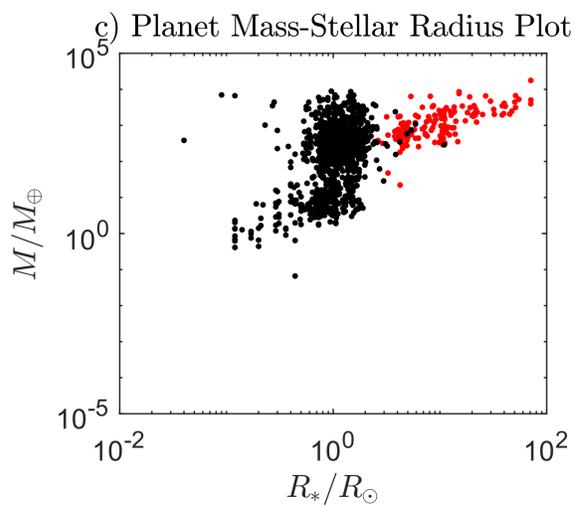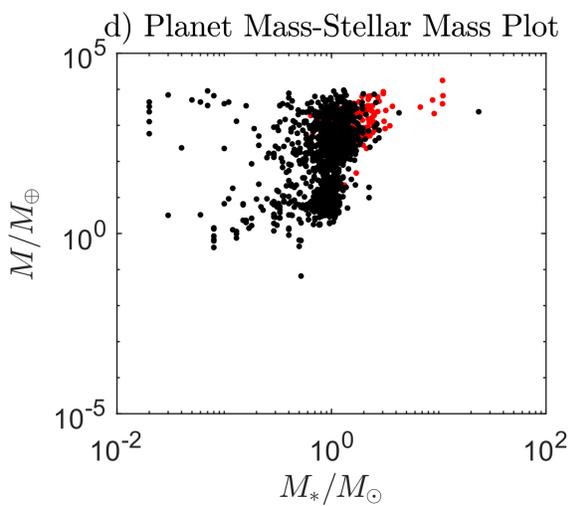